\documentclass{cs19proc}

\usepackage{hyperref}
\newcommand{\msun}{\mbox{\small{$\mathrm{M_{\odot}}$}}}		
\newcommand{\ltilde}{\mbox{\small{$\lesssim$}}}                                 
\newcommand{\less}{\mbox{\small{$<$}}}                                               
\newcommand{\plusminus}{\mbox{\small}{$\pm$}}                               
\newcommand{\equal}{\mbox{\small}{$=$}}                                            

\editors{G.~A. Feiden}
\publisher{Zenodo}
\conference{The 19th Cambridge Workshop on Cool Stars, Stellar Systems, and the Sun}
\conferencedate{2016}

\title{Clusters as benchmarks for measuring fundamental stellar parameters}
\author{Cameron P. M. Bell$^{1}$}

\affiliation{$^{1}$Institute for Astronomy, ETH Z{\"u}rich, 8093 Z{\"u}rich, Switzerland}

\shorttitle{Clusters}
\shortauthors{Cameron P. M. Bell}

\abs{In this contribution I will discuss fundamental stellar parameters as determined from young star clusters; specifically those with ages less than or approximately equal to that of the Pleiades. I will focus primarily on the use of stellar evolutionary models to determine the ages and masses of stars, as well as discuss the limitations of such models using a combination of both young clusters and eclipsing binary systems. In addition, I will also highlight a few interesting recent results from large on-going spectroscopic surveys (specifically \emph{Gaia}-ESO and APOGEE/IN-SYNC) which are continuing to challenge our understanding of the formation and early evolutionary stages of young clusters.}

\begin{document}

\maketitle

\section{Introduction}
\label{introduction}

This contribution constitutes but a small part of the two-day splinter session entitled ``Star clusters from space, from the ground, and over time'' which took place at the 19th Cambridge Workshop on Cool Stars, Stellar Systems, and the Sun in Uppsala, Sweden in June 2016. Given the nature of this contribution it will be far from comprehensive, both in terms of the range of fundamental stellar parameters covered, but also in terms of the various methods employed to estimate such parameters (for more information on these the reader is referred to other Cool Stars 19 proceedings contributions). A recent and more comprehensive discussion on star clusters which not only covers global properties of both young and old clusters, but also our current understanding of the dynamical evolution of clusters, can be found in the proceedings of the Ecole Evry Schatzman 2015 (EES2015) school ``Stellar clusters: benchmarks of stellar physics and galactic evolution''.

Clusters have long represented benchmarks with regard to the determination of fundamental stellar parameters, in large part due to the underlying assumption that members within such ensembles share several common properties; namely they are coeval, have the same chemical composition and are located at roughly the same distance. In addition to these shared characteristics, it is also observationally advantageous to focus on clusters as they have a significantly higher stellar number density (per unit area on the sky) compared to either young associations/moving groups or field stars and so for a given allocation of telescope time one can thus maximise the number of stars in ones sample.

Studies of clusters have also been instrumental in driving our understanding of the formation and evolution of stars. By studying a given cluster we can infer the mass dependence of astrophysical phenomena at a given epoch and by studying several clusters spanning a range of ages we can track how such phenomena evolve with time, as well as investigate second-order effects such as the local environment.

\section{Global parameters}
\label{cmd}

\begin{figure}[h]
\centering
\includegraphics[width=\columnwidth]{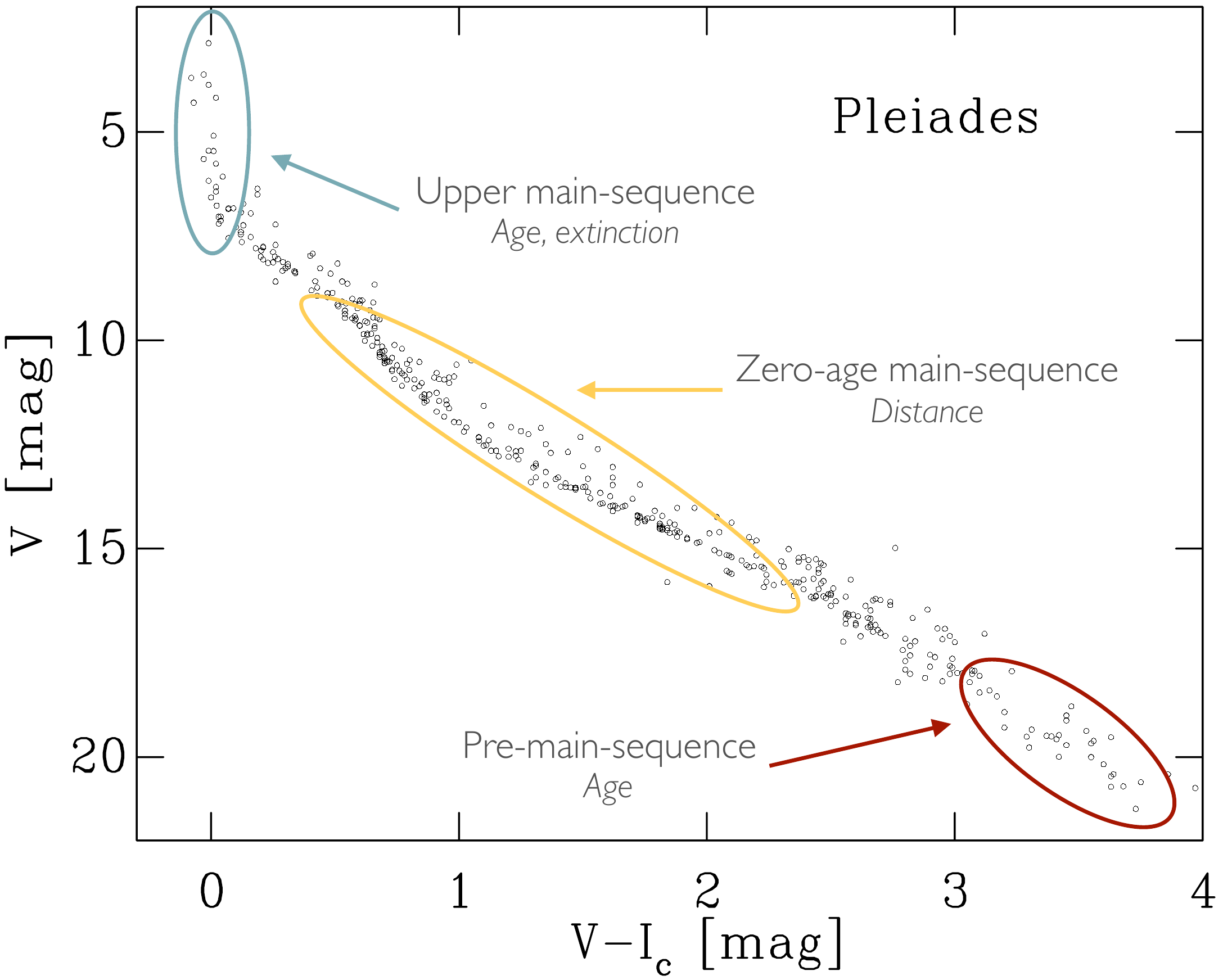} 
 \caption{The $V, V$--$I_{\rm{c}}$ CMD of the Pleiades with several regions marked and the global parameters one can estimate using stars in these regions.}
   \label{fig:pleiades_cmd_label}
\end{figure}

Arguably one of the simplest things to do with a cluster is to perform a multi-band photometric survey and create a colour-magnitude diagram (CMD) by plotting the magnitude versus the difference in magnitude in two different bandpasses. Fig.~\ref{fig:pleiades_cmd_label} shows the $V, V$--$I_{\rm{c}}$ CMD of the Pleiades with data taken from \cite{Stauffer07} in which several regions have been labelled as well as the global parameters one can estimate from stars in these regions. The Pleiades is arguably the best-studied (young) cluster with a rich sequence running from mid B-type stars down to the stellar/substellar boundary and so provides the ideal illustrative tool for the purposes of discussing the derivation of fundamental stellar parameters from clusters.

\subsection{Zero-age main-sequence}
\label{zams_distance}

At the age of the Pleiades the majority of stars have already settled onto the zero-age main-sequence (ZAMS) where they are steadily burning hydrogen in their cores. The ZAMS represents a temporally stable regime in which the stars essentially remain stationary with age in the CMD, and so whilst this region is not suitable for age estimates, it can however provide an estimate of the distance to the cluster; a prerequisite if one is interested in the bolometric luminosity of cluster members. So-called main-sequence fitting distances can be determined by comparing the sequence in question with either an empirical main-sequence relation based on photometry of nearby stars with known distances or via the use of theoretical stellar evolutionary models. There are of course issues related to both methods. For instance, empirical relations are typically mean relations computed from stars with a range of ages (of up to several Gyr) and so may not accurately reflect the positions of ZAMS stars in the CMD. The empirical sequence can be significantly brighter due to evolutionary effects and so will affect the apparent distance to the cluster (see e.g. \citealp{Littlefair10}). Similarly, stellar evolutionary models are not only susceptible to uncertainties in the underlying physics included, but must also be transformed from the theoretical Hertzsprung-Russell (H-R) plane to the observational CMD plane which requires the use of both a colour-$T_{\rm{eff}}$ relation to set the abscissa and bolometric correction (BC)-$T_{\rm{eff}}$ relations to set the ordinate. Regardless which method one adopts to infer the distance to a given cluster, the potential effects of both interstellar extinction and compositional differences must first be accounted for as both will act to modify the positions of stars in the CMD relative to unreddened and/or lower/higher metallicity stars.

Although it may seem that main-sequence fitting distances will be consigned to history in the age of \emph{Gaia}, it is worth remembering that there remains an outstanding discrepancy with regards to the \emph{Hipparcos} distance to the Pleiades and those estimated via main-sequence fitting (e.g. \citealp{An07}; cf. 120\,pc with 135\,pc) as well as other complementary methods (see e.g. \citealp{Soderblom05,Melis14}). Hence, although \emph{Gaia} will likely resolve this discrepancy, main-sequence fitting distances will continue to provide a simple, yet effective sanity check with regards to estimating cluster distances.

\subsection{Upper main-sequence}
\label{upper_ms_age_extinction}

The upper main-sequence (upper-MS) represents a region of the CMD which is not temporally stable i.e. as a function of age the evolution of stars becomes noticeable and can therefore be used to provide an age determination. When a star reaches the ZAMS it begins fusing hydrogen into helium in its core. Over time, the helium content of the core will increase and this leads to the star moving both redward and brighter in the CMD. Although subtle, this progression of the sequence between the ZAMS and terminal-age main-sequence (TAMS; where core hydrogen fusion ends) is a robust age indicator and the combination of stellar evolutionary models and sophisticated fitting techniques can provide reasonably well-constrained ages with statistically meaningful uncertainties (see e.g. \citealp{Naylor09, Bell13}).

\begin{figure}[t]
\centering
\includegraphics[width=\columnwidth]{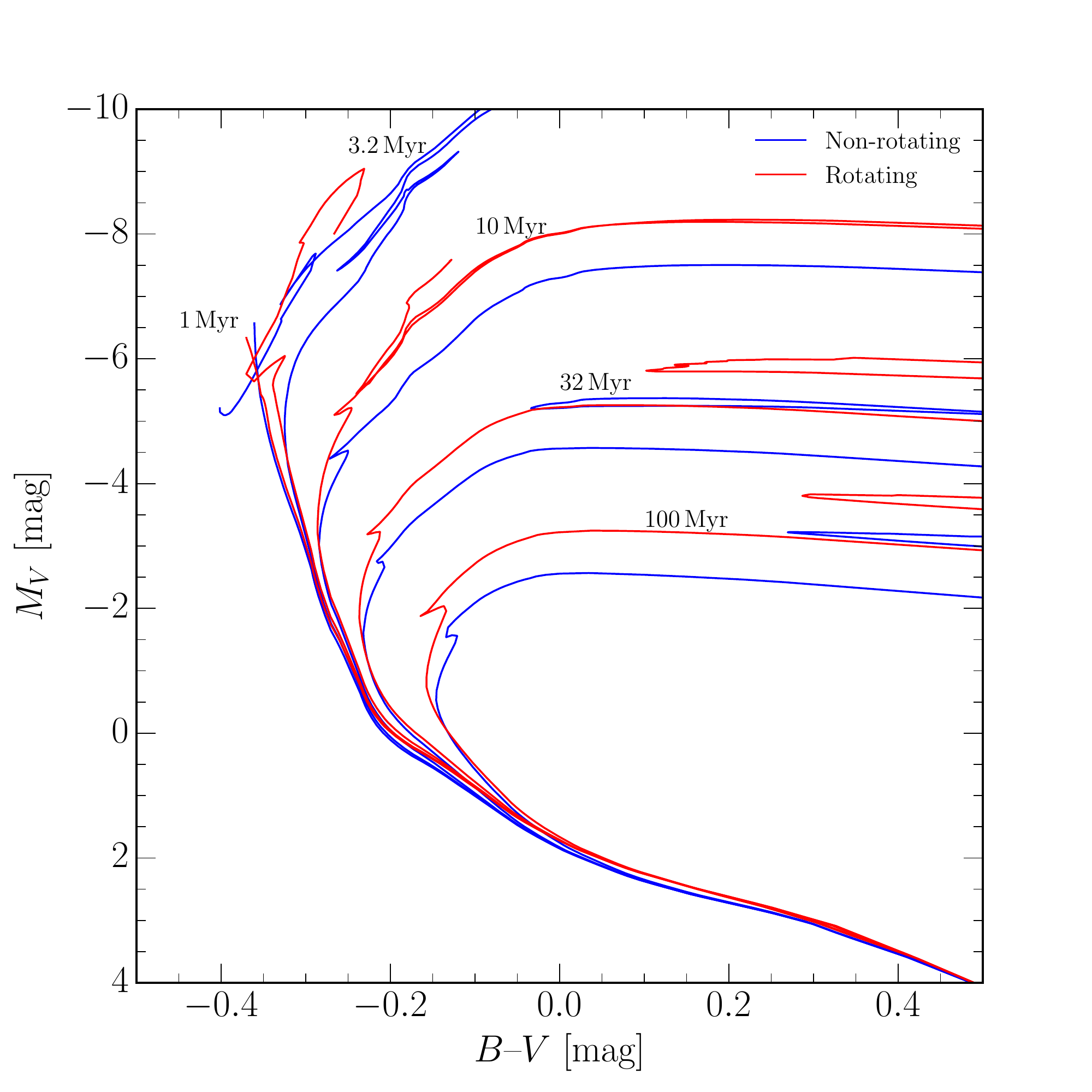} 
 \caption{The effects of including rotation in stellar evolutionary models in the $M_{V}, B$--$V$ CMD. The isochrones shown are from the recent Geneva models by \cite{Ekstrom12}.}
   \label{fig:upper}
\end{figure}

The use of stellar evolutionary models means that any age determination will naturally be model-dependent and as such is prone to uncertainties inherent in the models. In the high-mass regime the main sources of uncertainty are the effects of stellar rotation, the degree of convective core overshoot and, to a lesser extent, the rate of mass-loss in early-type stars. The inclusion of rotation and/or convective core overshooting will act to increase the main-sequence lifetime of a star as a result of increased levels of hydrogen supplied to the core (see e.g. \citealp{Meynet00}). Fig.~\ref{fig:upper} illustrates the effects of including stellar rotation in evolutionary models. The models used in this demonstration are the recent Geneva models by \cite{Ekstrom12} for which the authors assume a fixed rotation rate of 40\% of break-up. It is evident from the $M_{V}, B$--$V$ CMD that the inclusion of rotation (the only difference between the two sets of isochrones shown in Fig.~\ref{fig:upper}) makes a significant difference to the position of the isochrones in the CMD, and ultimately the derived age (see e.g. \citealp{Brandt15b}). For example, ages determined from the main-sequence turn-off would be affected at the $\sim$30\% level. On the other hand, the transition from the ZAMS to TAMS is much less affected -- and of course there are a greater number of stars in this region compared to the post-main-sequence due to the shape of the mass function -- and so age determinations from this region are affected at only the $\sim$10\% level. The main issue with main-sequence ages of course is the relative paucity of stars (especially in younger clusters) and so whilst the resultant age may have a moderately small systematic uncertainty, the statistical uncertainty can be significant.

In addition to age estimates, the higher mass stars (late B-type and earlier) can also be used to calculate the extinction towards a given cluster. This is typically performed using multi-band photometry blueward of (and including) the $V$-band in either a colour-colour diagram or through the use of various reddening vectors to de-redden individual sources onto a given sequence (e.g. the Q-method; \citealp{Johnson53}). This can trivially and quickly allow one to ascertain whether the extinction is uniform across the cluster or whether it is spatially variable. Note that not all young clusters/star-forming regions contain high-mass stars (see e.g. Taurus), and so in such cases both photometry and spectroscopy of individual members is necessary i.e. the combination of an observed colour and a spectral type will permit one to determine the extinction of a given star.

\subsection{Pre-main-sequence}
\label{ms_turn-on_age}

\begin{figure}[t]
\centering
\includegraphics[width=\columnwidth]{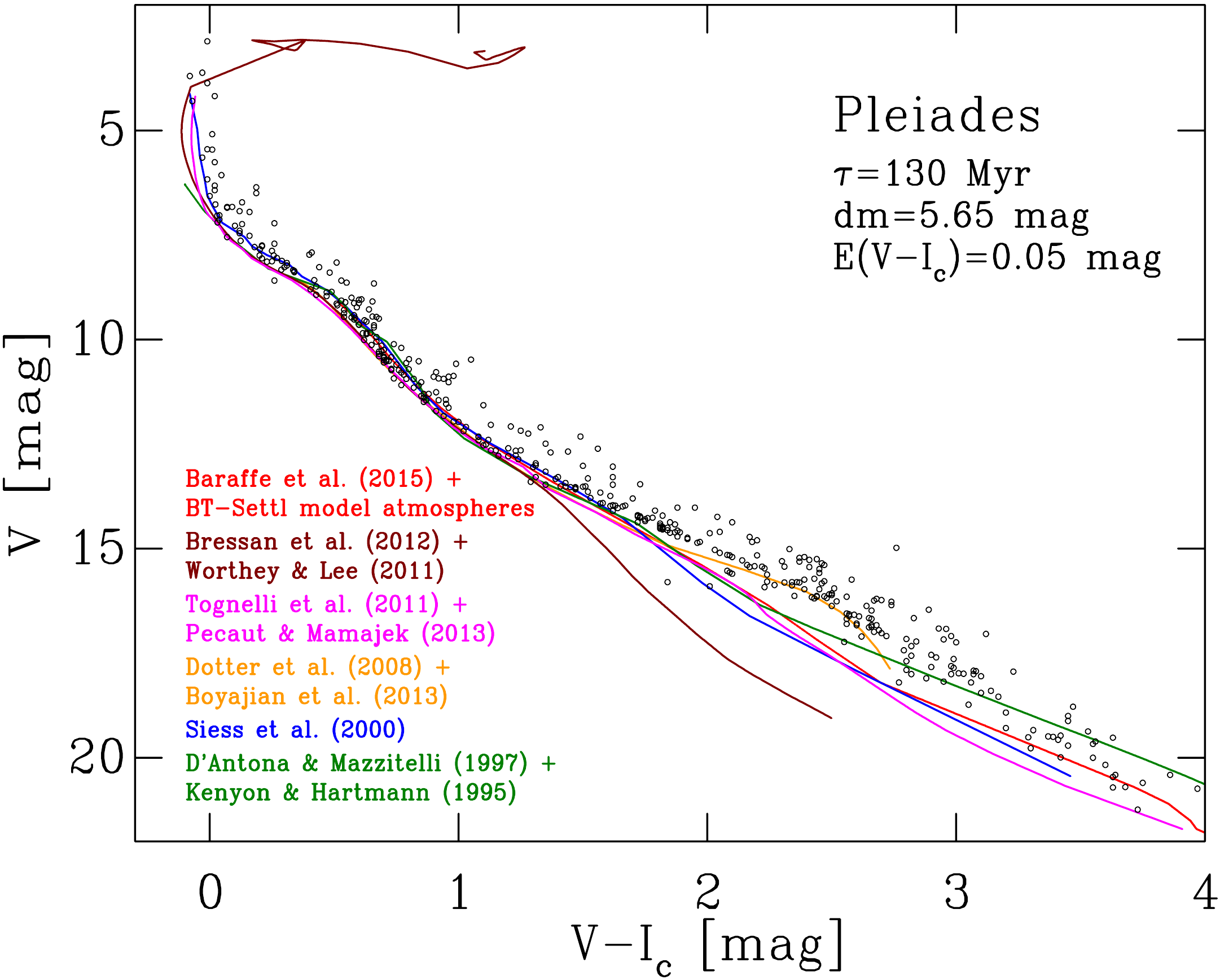} 
 \caption{The $V, V$--$I_{\rm{c}}$ CMD of the Pleiades with several sets of commonly used pre-MS isochrones overlaid. The upper reference refers to the stellar interior models, whereas the lower reference corresponds to the particular colour-$T_{\rm{eff}}$ and BC-$T_{\rm{eff}}$ relations used to transform the model into the observational plane.}
   \label{fig:pre-ms}
\end{figure}

Stars in the pre-main-sequence (pre-MS) phase are still contracting under the influence of gravity, and as such become noticeably fainter with age and hence provide an additional age diagnostic for a cluster. As with ages determined from the upper-MS, pre-MS ages are typically inferred via the use of stellar evolutionary models and are thus model-dependent. Unlike the upper-MS regime, however, there is a high degree of model dependency with pre-MS ages. Fig.~\ref{fig:pre-ms} illustrates this by showing several sets of publicly available pre-MS model isochrones which are commonly adopted in the literature overlaid on the $V, V$--$I_{\rm{c}}$ CMD of the Pleiades. The cluster parameters listed in the upper right of Fig.~\ref{fig:pre-ms} are all well-constrained and have been determined independently of fitting model isochrones in the CMD; namely the lithium depletion boundary (LDB) age from \cite{Barrado04b}, the VLBI distance of \cite{Melis14} and a reddening based on the mean extinction of \cite{Stauffer98a}. Fig.~\ref{fig:pre-ms} clearly shows a discrepancy between the observed Pleiades sequence in the CMD and the predicted colours/magnitudes of the models (having been transformed into the observational plane). Not only do none of the models match the sequence at cool temperatures (see e.g. \citealp{Bell12}), but furthermore they do so in a non-systematic way i.e. the derived age depends on which part of the sequence is fitted. Given the importance we place on the use of pre-MS model isochrones for estimating stellar ages, and by extension, timescales for important astrophysical phenomena in young (typically) less well-characterised clusters/associations, this is particularly perturbing.

If we are to use such models to infer ages from young star clusters it is apparent that some form of empirical correction to the BC-$T_{\rm{eff}}$ relation is required so as to fit an observed sequence in the CMD at a given age. Furthermore, if one wishes to use such ``semi-empirical'' models at significantly younger ages then any correction should include some form of surface gravity (log$\,g$) dependence (see e.g. \citealp{Bell14}). Note that even if such a correction is applied, the resultant ages from pre-MS models will still be heavily model-dependent and can differ by factors of 2--3 due to differences in the underlying input physics and parameters adopted in the models.

\subsection{Lithium depletion boundary}
\label{lithium_depletion_boundary}

\begin{figure}[t]
\centering
\includegraphics[width=\columnwidth]{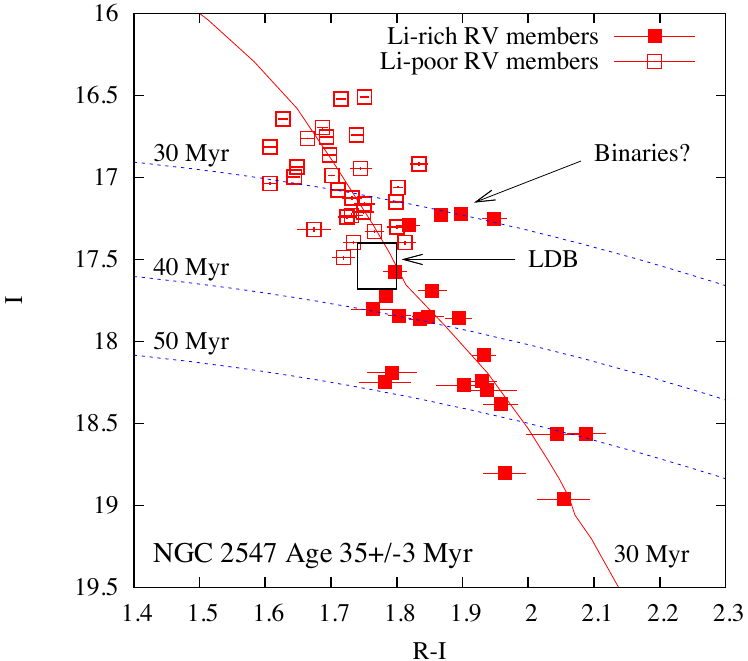} 
 \caption{The LDB in NGC~2547 defined by the clear discontinuity in the sequence between Li-poor and Li-rich members. The horizontal dashed lines denote constant luminosity loci with ages of 30, 40 and 50\,Myr. Figure from \protect\cite{Soderblom14}.}
   \label{fig:ldb}
\end{figure}

\begin{figure*}[t]
\centering
\includegraphics[width=\textwidth]{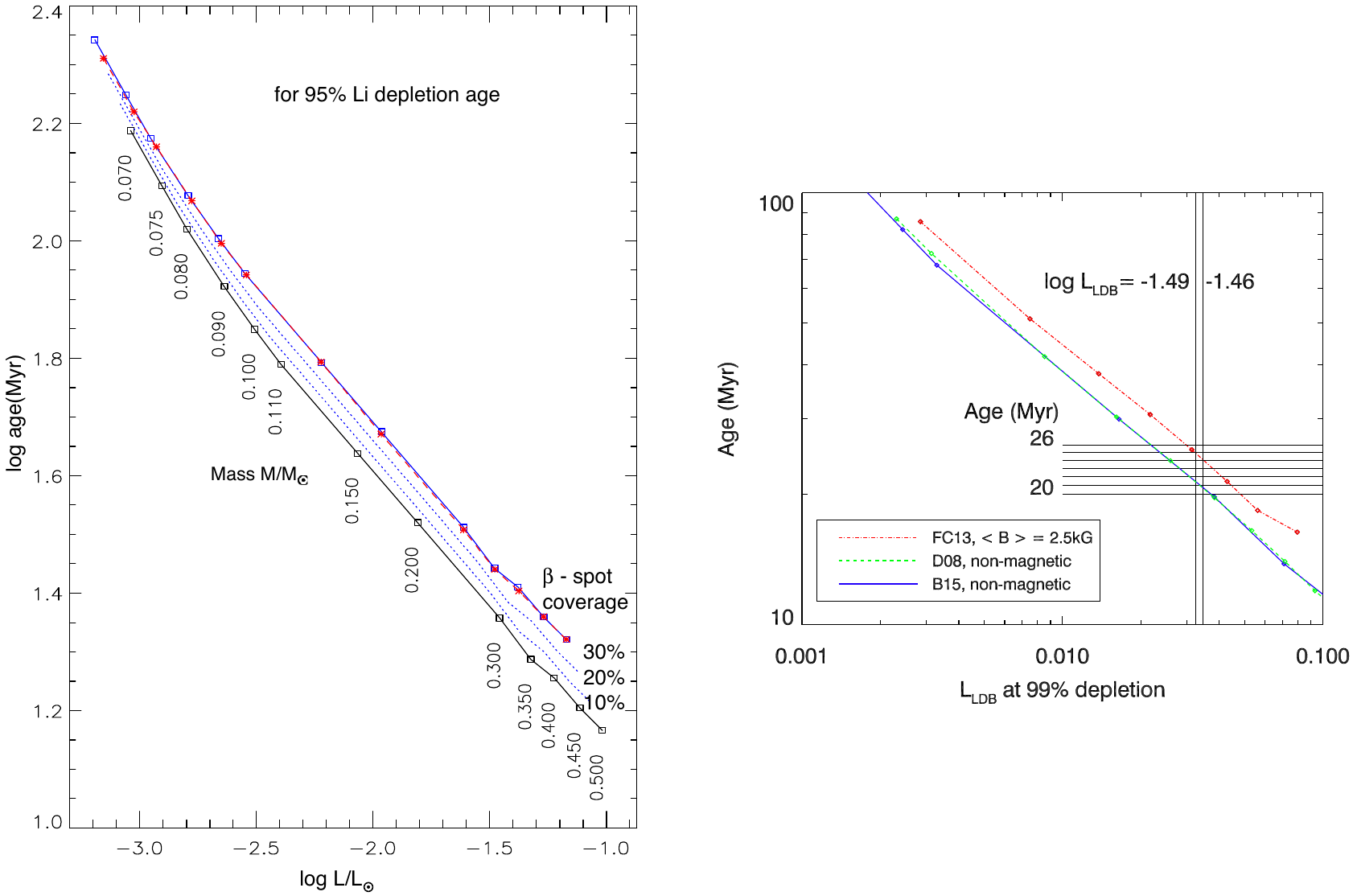} 
 \caption{The effects of including magnetic field-related phenomena on LDB ages. \emph{Left:} The effects of different star spot filling factors. Figure from \protect\cite{Jackson14b}. \emph{Right:} The effects of inhibiting convective flows. Figure from \protect\cite{Binks16}.}
   \label{fig:jackson_binks_ldb}
\end{figure*}

The basic premise underlying the LDB technique is that low-mass pre-MS stars (\ltilde\,0.4\,\msun) remain fully convective until they reach the ZAMS. As they contract, the core temperature increases until it is capable of burning Li at temperature of $\sim$\,3$\times$10$^{6}$\,K. Given the fully convective nature of the stars, Li-depleted material at the centre is brought up to the surface and Li-rich material from the surface brought to the centre where it becomes depleted, and complete Li depletion occurs in a fraction of a Kelvin-Helmholtz timescale. Fig.~\ref{fig:ldb} illustrates this rapid phase of Li destruction and the resultant sharp discontinuity in the $I, R$--$I$ CMD of NGC~2547 between stars which have contracted sufficiently to burn Li and those at slightly lower masses which have not. The LDB technique is evidently more telescope time intensive in the sense that one needs to first isolate potential members based on their positions in the CMD and then follow-up spectroscopy is required to identify the resonant Li feature at 6708\,\AA. The overriding benefit of this technique, however, is the high-degree of model-insensitivity. Unlike the ages derived from model isochrones in the CMD in which the various models simply do not agree, the different models agree remarkably well on the luminosity at which Li depletion occurs with systematic uncertainties at the $<$\,10\% level (see e.g. \citealp{Tognelli15}). Unfortunately, the LDB technique is practically only applicable in clusters with ages between 20 and 200\,Myr. At younger ages model dependency becomes increasingly problematic, reaching levels of $\sim$\,30\%, whereas at older ages the combined age and distance of potential target clusters means that the boundary itself simply becomes too faint. To date there are 10 clusters with LDB ages in the literature and these represent the necessary fiducial points against which other age-dating techniques can be validated in the same cluster (see e.g. \citealp{Mamajek14}).

\begin{figure*}[t]
\centering
\includegraphics[width=\textwidth]{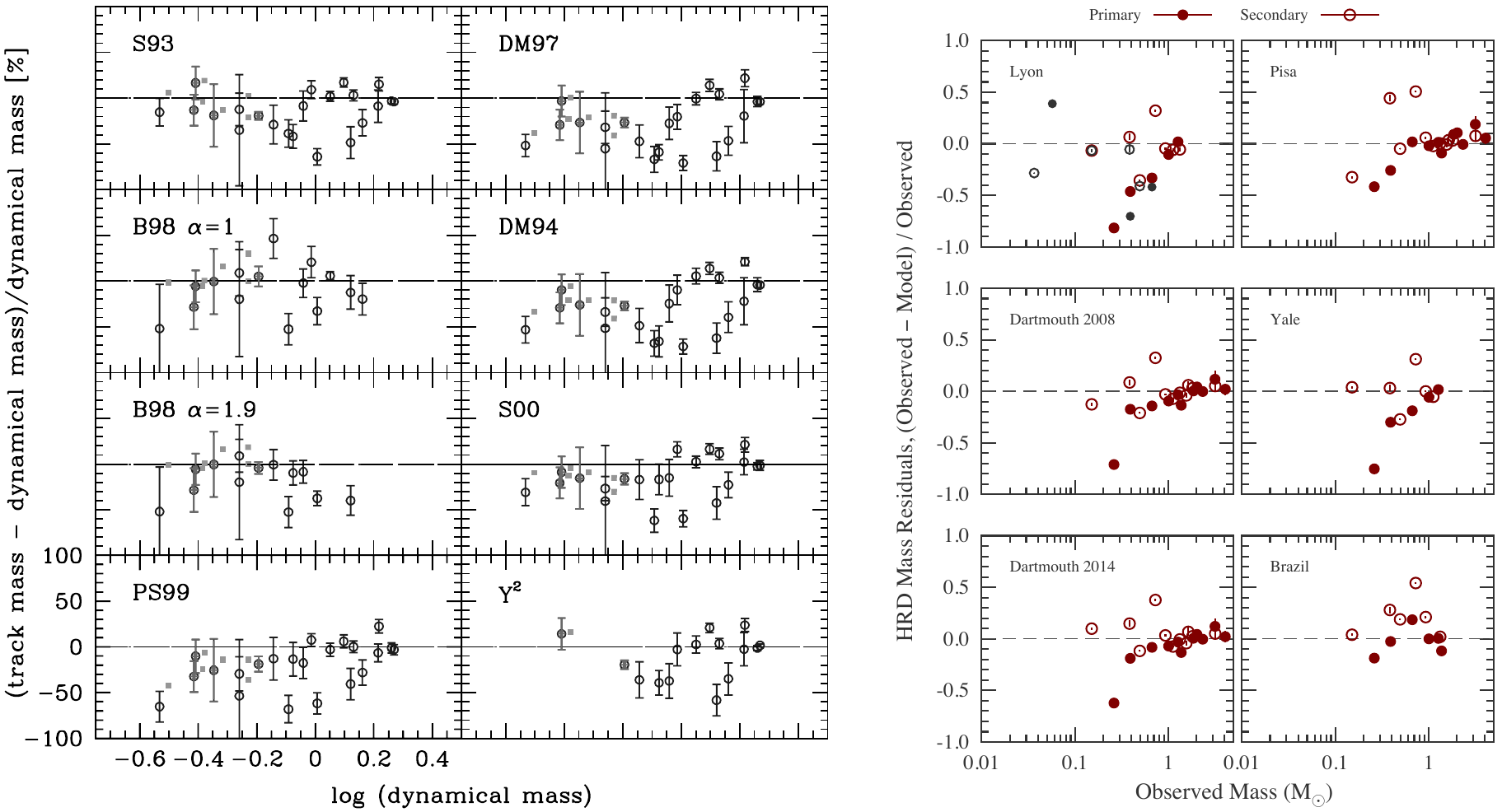} 
 \caption{\emph{Left:} Percentage difference between dynamically-determined stellar masses and those predicted by stellar evolutionary models for pre-MS stars in the H-R diagram. Figure from \protect\cite{Hillenbrand04}. \emph{Right:} Same as left-hand panel but showing only young eclipsing binaries. Figure from \protect\cite{Stassun14}.}
   \label{fig:hillenbrand_stassun_masses}
\end{figure*}

\subsection{The effects of magnetic fields}
\label{effects_magnetic_fields}

Although there is a very strong consensus amongst the various stellar evolutionary models as regards the luminosity of the onset of Li depletion, such models are well-known to be over simplistic representations of actual stars and do not include certain phenomena which we know to exist/occur in low-mass stars. Over the past couple of years several investigations have looked into the effects of including magnetic field-related phenomena in evolutionary models; namely the presence of star spots in the stellar photosphere and the inhibition of convective flows due to strong large-scale magnetic fields (see e.g. \citealp{Feiden14,Jackson14b,Somers15}). Both of these act to block outward flux from the star and hence slow the rate of contraction resulting in magnetic stars being cooler and larger at a given age than non-magnetic counterparts. As magnetic stars are cooler, it therefore takes longer to reach the necessary temperature in the core to burn Li and so the Li depletion timescale is extended. Fig.~\ref{fig:jackson_binks_ldb} illustrates the effects of including each of the aforementioned phenomena in stellar evolutionary models. Assuming a typical star spot coverage of $\sim$30\% (shown in the left-hand panel of Fig.~\ref{fig:jackson_binks_ldb}) or magnetic field strengths of a few kG (shown in the right-hand panel of Fig.~\ref{fig:jackson_binks_ldb}), the LDB age can increase by up to $\sim$20\% thereby suggesting that our current estimates represent lower limits to the cluster ages. Note the agreement between the two non-magnetic stellar evolutionary models in the right-hand panel of Fig.~\ref{fig:jackson_binks_ldb} which have been computed using different input physics and assumptions, that reiterates the statement concerning the high-level of model independence discussed in Section~\ref{lithium_depletion_boundary}.

\section{Young binary stars}
\label{binary_stars}

Binary stars represent a subset of stars which are sometimes referred to as ``benchmark'' or ``touchstone'' stars, in the sense that they permit tests of stellar evolutionary models at the most fundamental level by providing direct measurements of masses, radii, $T_{\rm{eff}}$ and bolometric luminosities. Of these properties, perhaps mass is of the greatest interest given that this is the primary input for stellar evolutionary models and essentially dictates the subsequent evolution of the star. Over the past decade several studies have collated the available dynamical information from young binary systems to test stellar evolutionary models, and in each case the same conclusion was reached; namely that modern evolutionary models are unable to accurately predict the properties of young, low-mass stars in binary systems (see e.g. \citealp{Hillenbrand04,Mathieu07,Stassun14}). The left-hand panel of Fig.~\ref{fig:hillenbrand_stassun_masses} shows that masses determined from positions in the H-R diagram are systematically underestimated for a given stellar $T_{\rm{eff}}$ and bolometric luminosity (i.e. radius). The study of \cite{Hillenbrand04}, however, was predominantly based on astrometric binaries with few eclipsing binaries in the sample. Over the past few years several such systems have been identified which led to a more recent study by \cite{Stassun14} who compared a newer generation of stellar evolutionary models to a selection of young eclipsing binaries only. Despite the inclusion of more up-to-date physics (e.g. opacities and molecular line lists), the models essentially fared no better than those investigated by \cite{Hillenbrand04}. For masses above 1\,\msun\ all models predict masses to within 10\% of the dynamical measurements, however below 1\,\msun\ the situation is much worse with mass errors of 50--100\%. Note a qualitative difference between the two panels in Fig.~\ref{fig:hillenbrand_stassun_masses}, namely that the \cite{Hillenbrand04} study found that the models tend to underestimate the mass, whereas the \cite{Stassun14} investigation found that the models overestimate the masses (with a potential dichotomy between the predicted primary and secondary masses). There are only three eclipsing binaries and one set of evolutionary models in common between the two studies for which very similar results were found.

\begin{figure*}[t]
\centering
\includegraphics[width=0.8\textwidth]{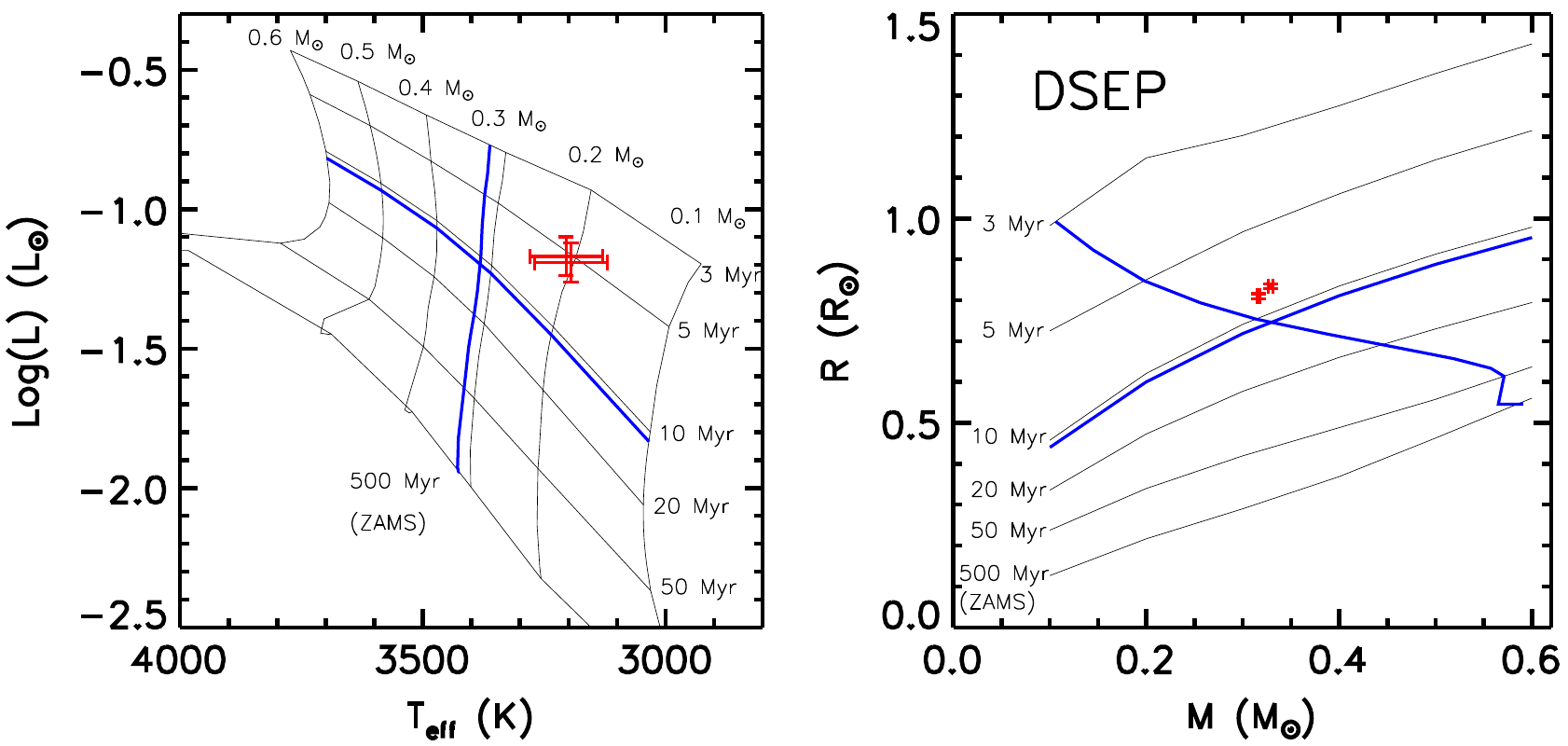} 
 \caption{\emph{Left:} The primary and secondary components of UScoCTIO 5 in the H-R diagram. The isochrones and mass tracks are from the Dartmouth stellar evolutionary models of \protect\cite{Dotter08}. The blue lines denote the expected age of the system and the dynamically-determined mass of both components. \emph{Right:} Both components in the mass-radius diagram. The blue lines represent the expected age of the system and the measured luminosity of both components. Figure from \protect\cite{Kraus15}.}
   \label{fig:kraus_binary}
\end{figure*}

Recently, \cite{Kraus15} identified UScoCTIO 5, a known spectroscopic binary with spectral type M4.5, as an eclipsing binary with both eclipses apparent in K2 light curves. Fig.~\ref{fig:kraus_binary} shows the primary and secondary components of UScoCTIO 5 (which are almost of identical mass) in both the H-R and mass-radius diagrams. The blue lines in both panels represent the expected age, dynamical mass and measured luminosity of both components. It is clear from Fig.~\ref{fig:kraus_binary} that despite the on-going uncertainty with regards to the age one should adopt for particular stars in Upper Scorpius (see \citealp{Pecaut16}) the stellar evolutionary models (in 4 out of 5 cases) underestimate the masses. The notable exception to this are the recent PARSEC models of \cite{Chen14} which include an empirical correction to the outer boundary condition based on observations of low-mass stars in significantly older clusters and which appears to be an over-correction in the case of UScoCTIO 5.

There is of course a fundamental issue with using the H-R diagram to test stellar evolutionary models, which simply arises from the difficulty of measuring both the $T_{\rm{eff}}$ and bolometric luminosity. The former requires the use of model atmospheres which may still suffer from uncertainties regarding the underlying physics (especially in terms of missing sources of opacity) and the latter necessitates accurate and precise distance estimates (although upcoming \emph{Gaia} data releases will certainly help in this area). Thus, an even more fundamental test of stellar evolutionary models is to use the directly measured masses and radii and perform the comparison in the mass-radius diagram. The right-hand panel of Fig.~\ref{fig:kraus_binary} demonstrates that the models predict radii which are too small for the known luminosity and mass of the stars, which subsequently implies that the model $T_{\rm{eff}}$ are too high. This $T_{\rm{eff}}$ offset could be due to a number of potential issues including underlying issues with the prescription of convection in low-mass stars, missing opacities as well as a miscalibration of the spectral type-$T_{\rm{eff}}$ scale.

\begin{figure*}[t]
\centering
\includegraphics[width=0.7\textwidth]{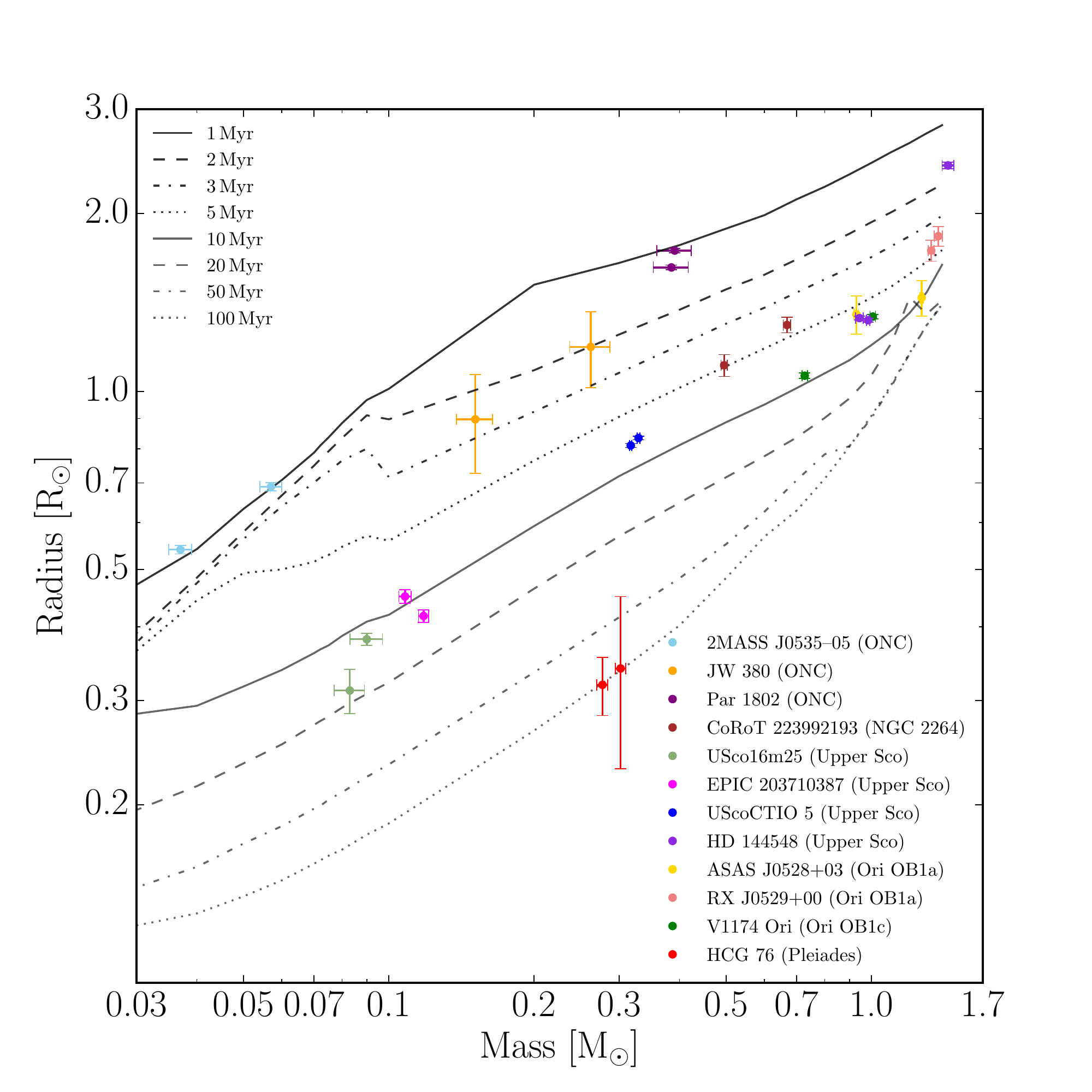} 
 \caption{The mass-radius diagram of young eclipsing binaries in which both the primary and secondary components have masses of \ltilde\,1.4\,\msun\ with the recent stellar evolutionary models of \cite{Baraffe15} overlaid.}
   \label{fig:binary}
\end{figure*}

Fig.~\ref{fig:binary} shows the mass-radius diagram for eclipsing binaries with ages of the Pleiades and younger in which both the primary and secondary components have masses of \ltilde\,1.4\,\msun, and which are compared against the recent stellar evolutionary models of \cite{Baraffe15}. There are a couple of points worth noting. First, although the number of systems has increased over the past few years, there is still a paucity of young eclipsing binaries with which to rigorously test evolutionary models. It is likely that the continuing K2 mission will help identify more such systems in the coming years. For example, if one can identify a number of young eclipsing binaries spanning a range of masses at a given age, across a range of ages, this will then not only permit stringent tests of the models, but may also perhaps allow us to constrain the uncertain physics inherent to them. Second, there appears to be a mixture of binary systems which are coeval and systems which are not. Such instances have previously been discussed in the literature (see e.g. \citealp{Gennaro12}), however it is worth noting that even the comparison of data and models in the mass-radius diagram is not strictly a fair comparison. For example, one is comparing low-mass, likely magnetically active stars in a binary system (with an increased potential for strong dynamical interactions at the earliest phases) with standard single-star, non-magnetic models. This could further be complicated by possible episodic accretion histories which may have dramatic effects on the internal structure of the stars (see e.g. \citealp{Audard14}).

\section{Spectroscopic surveys}
\label{spectroscopic_surveys}

\begin{figure*}[t]
\centering
\includegraphics[width=\textwidth]{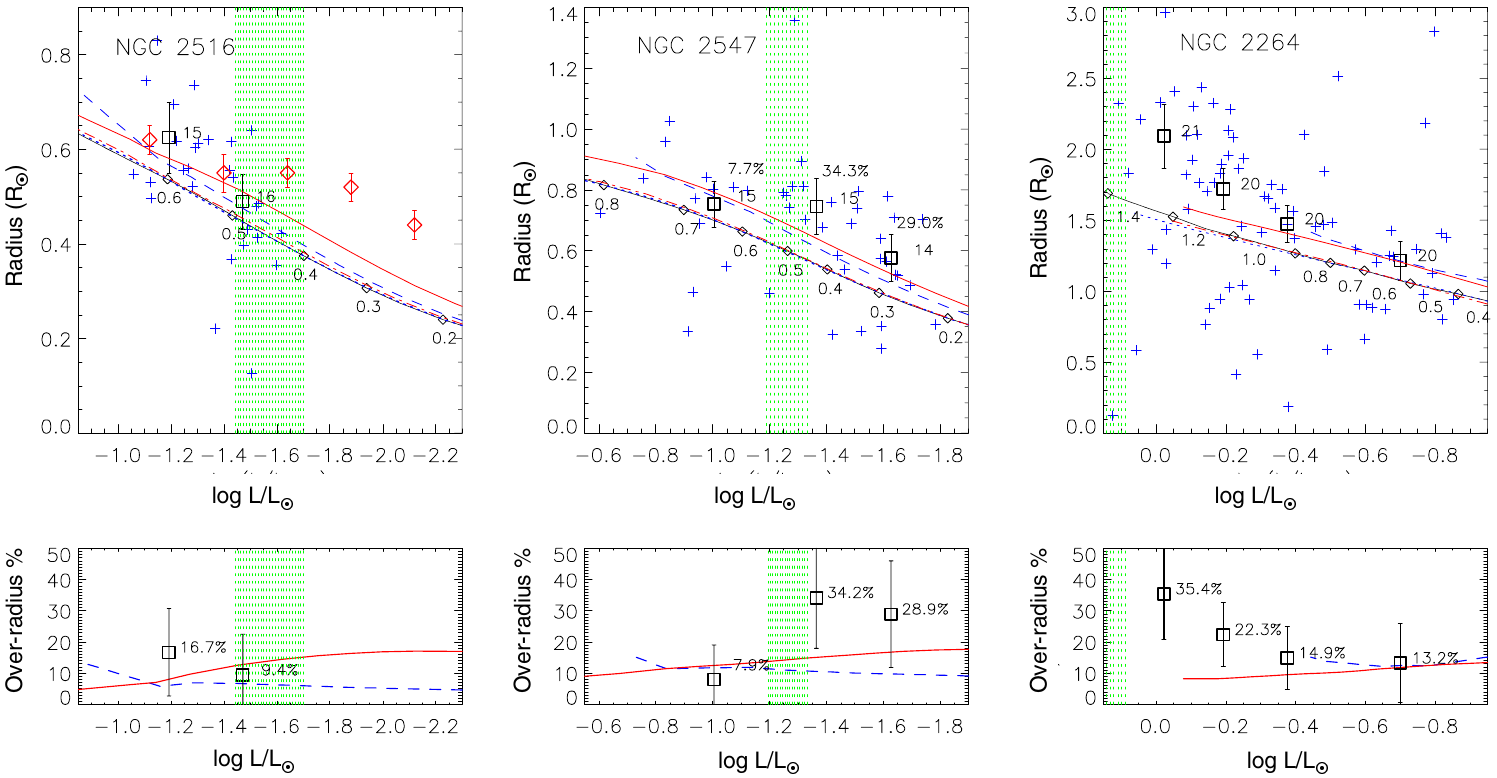} 
 \caption{\emph{Top:} Stellar radius versus bolometric luminosity for stars in three young clusters; NGC~2516 (140\,Myr), NGC~2547 (35\,Myr) and NGC~2264 (5\,Myr). The black squares denote the averaged radii in specific luminosity bins. The following stellar evolutionary models are shown for comparison: BHAC15 (black solid; \citealp{Baraffe15}), Dartmouth (blue dotted; \citealp{Dotter08}), Dartmouth modified to include magnetic fields (blue dashed; \citealp{Feiden15}), YREC (red dot-dashed; \citealp{Demarque08}), and YREC modified with an effective star spot coverage of 30\% (red solid; \citealp{Somers15}). The green shaded region represents the expected luminosities over which stars develop radiative cores according to the BHAC15 models; stars to the left have radiative cores, whereas those to the right are fully convective. \emph{Bottom:} The percentage difference between the measured radii and those predicted by the BHAC15 models. The blue dashed and red solid lines represent the percentage difference between the non-magnetic and modified Dartmouth and YREC evolutionary models respectively. Figure from \protect\cite{Jackson16}.}
   \label{fig:jackson_radii}
\end{figure*}

In the last few years the \emph{Gaia}-ESO and APOGEE/IN-SYNC surveys have collected large-scale spectroscopic samples of stars in young clusters. The primary benefit of such surveys is the homogeneous datasets which have been produced for the community which permit the investigation of common features and peculiarities of different clusters in a self-consistent and standardised way. This represents a marked shift from bringing together the hodgepodge of smaller surveys by different groups looking at different subsets of stars in a cluster and which have not only been collected using a variety of instruments, but have also been reduced and analysed in a heterogeneous way.

\begin{figure*}[hbtp!]
\centering
\includegraphics[width=\textwidth]{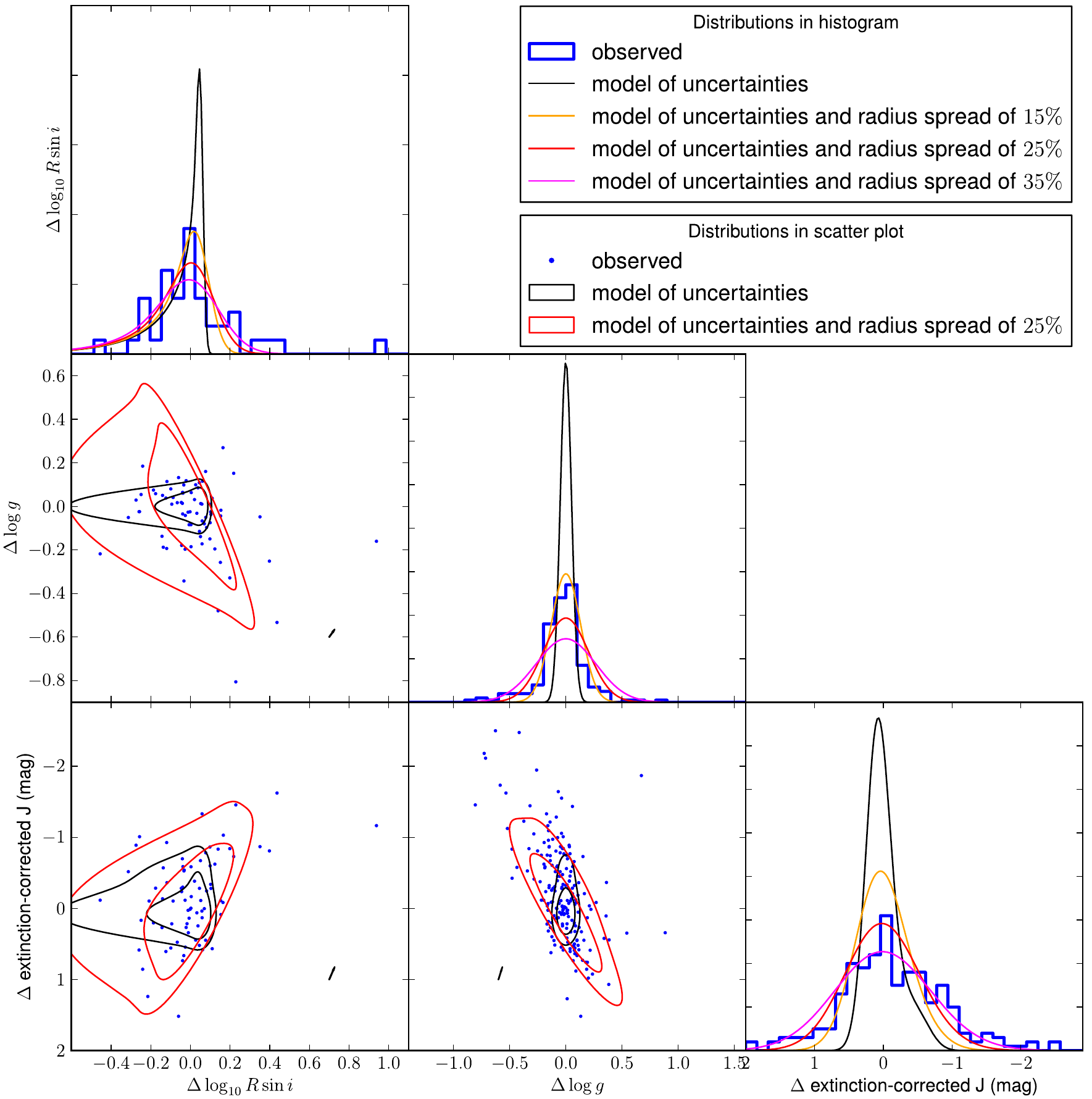} 
 \caption{Distributions of three (almost) independent stellar radius estimates for members of IC~348. The blue points and histograms show the observed distributions. The black distributions represent the expected distributions accounting for observational effects which could introduce an observed spread in stellar radii including binarity, measurement uncertainties and projection effects. The coloured distributions are the same as the black distributions but also include a Gaussian intrinsic stellar radius spread with a width of 15\% (orange), 25\% (red) and 35\% (magenta) around the median stellar radius at the corresponding $T_{\rm{eff}}$. Figure from \protect\cite{Cottaar14}.}
   \label{fig:cottaar_ic348}
\end{figure*}

\subsection{Inflated radii and radii spreads}
\label{inflated_radii_radii_spreads}

Observations have long suggested that the radii of short-period eclipsing binaries are inflated with respect to those predicted by standard stellar evolutionary models (see e.g. \citealp{Kraus11}). Combining rotational periods with projected equatorial velocities from the \emph{Gaia}-ESO survey, \cite{Jackson16} estimated the average radii of cluster members as a function of luminosity and age to investigate whether similarly inflated radii are observed in young, low-mass, rapidly-rotating stars. Fig.~\ref{fig:jackson_radii} shows the average radii of cluster members as a function of luminosity for three young clusters; NGC~2516 (140\,Myr), NGC~2547 (35\,Myr) and NGC~2264 (5\,Myr), which are compared against several sets of stellar evolutionary models (including modified versions of both the Dartmouth and YREC models which incorporate a prescription for magnetic fields and a star spot coverage of 30\% respectively, see \citealp{Feiden15,Somers15}). The lower panels of Fig.~\ref{fig:jackson_radii} display the percentage difference between the observed and predicted radii from the BHAC15 models \citep{Baraffe15}, whereas the blue dashed and red solid lines represent the percentage difference between the non-magnetic and modified Dartmouth and YREC evolutionary models respectively. In short, the radii of young cluster members are larger than those predicted by standard stellar evolutionary models and is more pronounced in fully convective low-mass stars in which it can reach levels of $\sim$30\%. With the modified evolutionary models currently available, it appears as though strong surface magnetic fields (exceeding 2.5\,kG), star spots covering $\sim$30\% of the photosphere or, more likely, a combination of both is necessary to explain the observed data.

The \emph{Gaia}-ESO and IN-SYNC spectroscopic surveys are providing high-precision fundamental stellar parameters (e.g. $\sigma_{\rm{RV}}$\,\ltilde\,0.2\,km\,s$^{-1}$, $\sigma_{\mathrm{log}\,g}$\,\less\,0.1\,dex, $\sigma_{T_{\rm{eff}}}$\,\less\,100\,K) and such precision is allowing us to investigate the constituent members of young star clusters in unprecedented detail. The first cluster observed within the framework of the APOGEE/IN-SYNC ancillary programme was IC~348 (see \citealp{Cottaar14}). In brief, the IN-SYNC programme has acquired thousands of high-resolution $H$-band spectra for thousands of pre-MS stars in young clusters. From these spectra they extract various parameters including $T_{\rm{eff}}$, surface gravity (log\,$g$) and rotational velocities ($v$ sin\,$i$) via comparison with stellar atmospheric models. Fig.~\ref{fig:cottaar_ic348} shows the observed distributions for three (almost) independent stellar radius estimates for members of IC~348; namely the projected stellar radius, surface gravity and extinction-corrected $J$-band magnitude. From Fig.~\ref{fig:cottaar_ic348} it is clear that compared to the expected distributions, which account for observational effects that could introduce an observed radius spread including binarity, measurement uncertainties and projection effects, the observed distributions are significantly wider. \cite{Cottaar14} find a best-fit to the observed distributions by combining this expected distribution with a Gaussian intrinsic stellar radius spread of width 25\% around the median stellar radius at the corresponding $T_{\rm{eff}}$. Furthermore, \citeauthor{Cottaar14} argue that all three radius diagnostics are correlated with the brighter stars tending to have significantly lower surface gravities and larger projected stellar radii, and that uncertainties in the derived parameters are not responsible for the observed spread.

The simplest explanation for an intrinsic spread in stellar radii in IC~348 is that this is a result of an intrinsic age spread within the cluster i.e. the younger stars have not contracted as much as the older stars in the same cluster and hence have correspondingly lower surface gravities and larger projected stellar radii. The exact age spread, however, is dependent upon the mean age of the IC~348 which lies between 3 and 6\,Myr. The main uncertainty in the age of the cluster is the uncertainty on its distance (ranging from 220 to 350\,pc; see e.g. \citealp{Herbst08}) and so hopefully this is another instance in which \emph{Gaia} will provide relief in the coming years. Assuming the older age of 6\,Myr, this corresponds to an upper limit on the age spread of 8\,Myr. Note, however, that \cite{Cottaar14} also find that the more rapid rotators have larger stellar radii and thus it is likely that the spread in stellar radii is not solely due to an intrinsic age spread, but may also include contributions from different accretion histories and/or different levels of magnetic activity amongst cluster members. The ability to disentangle these potential effects from one another to infer any genuine intrinsic age spread is currently beyond our means, however if intrinsic age spreads of a few Myr are common amongst the youngest clusters/star-forming regions (see also \citealp{DaRio16}) then this could have serious implications, especially regarding their use as fiducial age ``points'' in understanding, for example, circumstellar disc dissipation timescales, and hence planet formation timescales, as by definition the cluster does not have a single age, but a range.

\subsection{Kinematic substructure}
\label{kinematic_substructure}

\begin{figure}[t]
\centering
\includegraphics[width=0.45\textwidth]{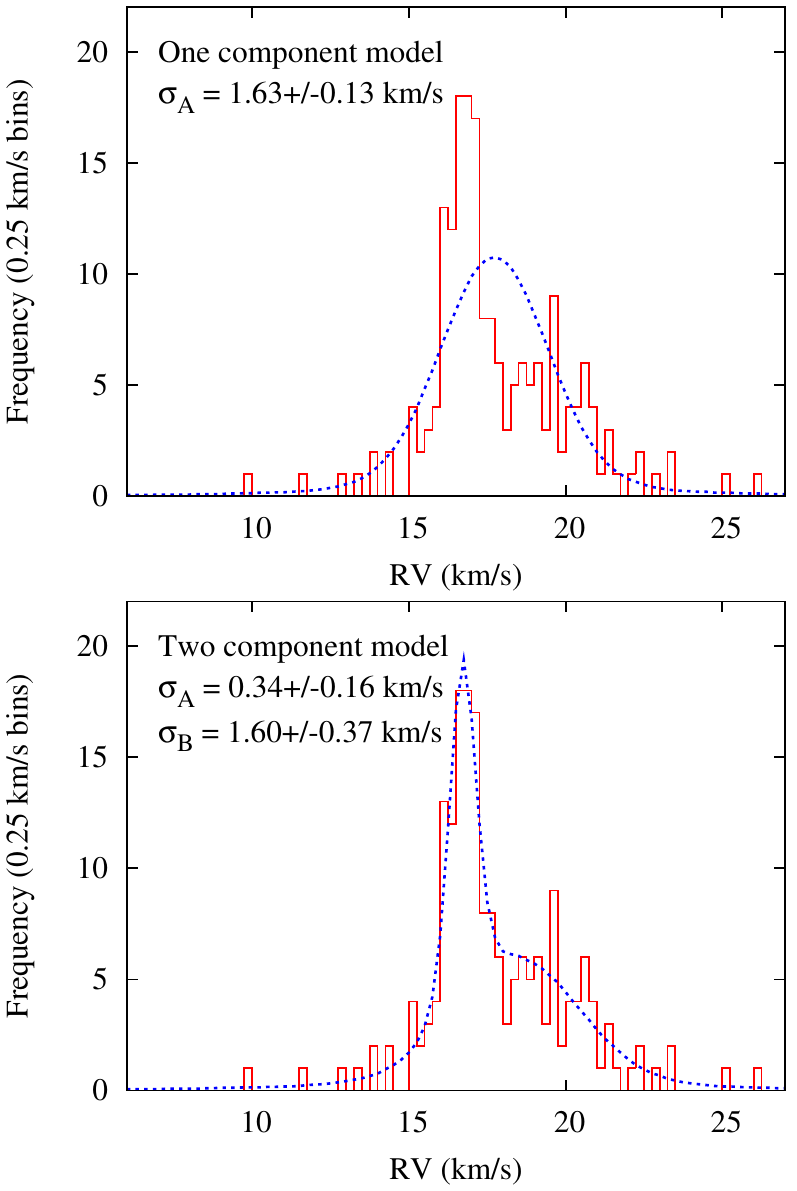} 
 \caption{Binned radial velocity histogram for members of the $\gamma^{2}$ Velorum cluster. \emph{Top:} Best-fit for a model consisting of a single Gaussian component including unresolved binaries. \emph{Bottom:} Same as the upper panel but instead the model consists of two Gaussian components each with an unresolved binary population. Figure from \protect\cite{Jeffries14}.}
   \label{fig:jeffries_gamma_vel}
\end{figure}

The low-mass stellar population associated with the Wolf-Rayet binary $\gamma^{2}$ Velorum was the first young ``cluster'' target observed as part of the \emph{Gaia}-ESO survey (see \citealp{Jeffries14}). Fig.~\ref{fig:jeffries_gamma_vel} shows the binned radial velocity histogram for $\gamma^{2}$ Velorum members. It is apparent that the best-fit model consisting of a single Gaussian component, which includes a fraction of unresolved binaries according to the results of the recent \citealp{Raghavan10} survey, represents a poor fit to the data. By contrast, the best-fit model comprised of two Gaussian components, each with an unresolved binary population, is a much better fit. The small uncertainties on the measured radial velocities are such that \citeauthor{Jeffries14} were able to unambiguously identify two distinct kinematic populations (hereafter referred to as Populations A and B, each of which contains approximately equal numbers of stars); one with a very narrow intrinsic width and the other with a much broader width (cf. $\sigma_{\rm{RV, A}}$\,\equal\,\,0.34\,\plusminus\,0.16 and $\sigma_{\rm{RV, B}}$\,\equal\,1.60\,\plusminus\,0.37\,km\,s$^{-1}$) and which is offset from the first by 2\,km\,s$^{-1}$. \citeauthor{Jeffries14} also identified several other differences between the two Populations, notably: i) based on levels of Li depletion, Population~A is about 1--2\,Myr older than Population~B and ii) Population~A appears to be in viral equilibrium and likely represents the bound remnant of an initially larger cluster which formed in the denser region of the Vela~OB2 association, whereas Population~B comprises a dispersed population of unbound stars which probably formed in the less dense regions of the association. Interestingly, the low-mass stars of Populations~A and B appear to be several Myr older than $\gamma^{2}$ Velorum, thus suggesting a scenario in which it formed after the bulk of the low-mass population and which possibly resulted in the termination of star formation in the region via the expulsion of gas and dust.

\section{Summary of conclusions}
\label{concluding_thoughts}

Below I briefly reiterate the main conclusions from this contribution.

\begin{enumerate}
\item The colour-magnitude diagram of a given cluster can provide several global parameters shared by constituent members (including age, distance and the presence/uniformity of interstellar extinction), although one should be aware of the underlying uncertainties as regards the use of stellar evolutionary models and carefully assess the pedigree of adopted empirical relations.
\item Model-dependent estimates of low-mass pre-MS stellar parameters are unreliable. Specifically, below 1\,\msun\ the models tend to underestimate the mass of a given star based on its position in the H-R diagram as well as predict radii which are too small based on its position in the mass-radius diagram.
\item There is tentative evidence that the introduction of magnetic field-related phenomena (such as star spots and/or the inhibition of convective flows) may help to resolve the discrepancy between dynamically-determined parameters and those predicted by stellar evolutionary models.
\item Recent spectroscopic survey results have demonstrated that clusters are more complex entities than previously thought (e.g. age spreads and kinematic substructure) and the continuing \emph{Gaia}-ESO and APOGEE/IN-SYNC surveys will only highlight further examples of this and hence continue to shape our understanding of the formation and early evolutionary stages of young clusters.
\end{enumerate}

\section*{Acknowledgments}
{I would like to thank the organisers of the splinter session for inviting me to present this review as well as the SOC/LOC who together provided an excellent scientific programme and an exceptionally ran conference.}

\bibliographystyle{mn3e3}
\bibliography{references}

\begin{thebibliography}{41}
\expandafter\ifx\csname natexlab\endcsname\relax\def\natexlab#1{#1}\fi

\bibitem[{{An} {et~al.}(2007){An}, {Terndrup}, {Pinsonneault}, {Paulson},
  {Hanson}, \& {Stauffer}}]{An07}
{An}, D., {Terndrup}, D.M., {Pinsonneault}, M.H., {Paulson}, D.B., {Hanson},
  R.B. et~al, 2007, \apj, 655, 233

\bibitem[{{Audard} {et~al.}(2014){Audard}, {{\'A}brah{\'a}m}, {Dunham},
  {Green}, {Grosso}, {Hamaguchi}, {Kastner}, {K{\'o}sp{\'a}l}, {Lodato},
  {Romanova}, {Skinner}, {Vorobyov}, \& {Zhu}}]{Audard14}
{Audard}, M., {{\'A}brah{\'a}m}, P., {Dunham}, M.M., {Green}, J.D., {Grosso},
  N. et~al, 2014, Protostars and Planets VI, 387

\bibitem[{{Baraffe} {et~al.}(2015){Baraffe}, {Homeier}, {Allard}, \&
  {Chabrier}}]{Baraffe15}
{Baraffe}, I., {Homeier}, D., {Allard}, F. and {Chabrier}, G., 2015, \aap, 577,
  A42

\bibitem[{{Barrado y Navascu{\'e}s} {et~al.}(2004){Barrado y Navascu{\'e}s},
  {Stauffer}, \& {Jayawardhana}}]{Barrado04b}
{Barrado y Navascu{\'e}s}, D., {Stauffer}, J.R. and {Jayawardhana}, R., 2004,
  \apj, 614, 386

\bibitem[{{Bell} {et~al.}(2012){Bell}, {Naylor}, {Mayne}, {Jeffries}, \&
  {Littlefair}}]{Bell12}
{Bell}, C.P.M., {Naylor}, T., {Mayne}, N.J., {Jeffries}, R.D. and {Littlefair},
  S.P., 2012, \mnras, 424, 3178

\bibitem[{{Bell} {et~al.}(2013){Bell}, {Naylor}, {Mayne}, {Jeffries}, \&
  {Littlefair}}]{Bell13}
{Bell}, C.P.M., {Naylor}, T., {Mayne}, N.J., {Jeffries}, R.D. and {Littlefair},
  S.P., 2013, \mnras, 434, 806

\bibitem[{{Bell} {et~al.}(2014){Bell}, {Rees}, {Naylor}, {Mayne}, {Jeffries},
  {Mamajek}, \& {Rowe}}]{Bell14}
{Bell}, C.P.M., {Rees}, J.M., {Naylor}, T., {Mayne}, N.J., {Jeffries}, R.D.
  et~al, 2014, \mnras, 445, 3496

\bibitem[{{Binks} \& {Jeffries}(2016)}]{Binks16}
{Binks}, A.S. and {Jeffries}, R.D., 2016, \mnras, 455, 3345

\bibitem[{{Brandt} \& {Huang}(2015)}]{Brandt15b}
{Brandt}, T.D. and {Huang}, C.X., 2015, \apj, 807, 24

\bibitem[{{Chen} {et~al.}(2014){Chen}, {Girardi}, {Bressan}, {Marigo},
  {Barbieri}, \& {Kong}}]{Chen14}
{Chen}, Y., {Girardi}, L., {Bressan}, A., {Marigo}, P., {Barbieri}, M. et~al,
  2014, \mnras, 444, 2525

\bibitem[{{Cottaar} {et~al.}(2014){Cottaar}, {Covey}, {Meyer}, {Nidever},
  {Stassun}, {Foster}, {Tan}, {Chojnowski}, {da Rio}, {Flaherty}, {Frinchaboy},
  {Skrutskie}, {Majewski}, {Wilson}, \& {Zasowski}}]{Cottaar14}
{Cottaar}, M., {Covey}, K.R., {Meyer}, M.R., {Nidever}, D.L., {Stassun}, K.G.
  et~al, 2014, \apj, 794, 125

\bibitem[{{Da Rio} {et~al.}(2016){Da Rio}, {Tan}, {Covey}, {Cottaar}, {Foster},
  {Cullen}, {Tobin}, {Kim}, {Meyer}, {Nidever}, {Stassun}, {Chojnowski},
  {Flaherty}, {Majewski}, {Skrutskie}, {Zasowski}, \& {Pan}}]{DaRio16}
{Da Rio}, N., {Tan}, J.C., {Covey}, K.R., {Cottaar}, M., {Foster}, J.B. et~al,
  2016, \apj, 818, 59

\bibitem[{{Demarque} {et~al.}(2008){Demarque}, {Guenther}, {Li}, {Mazumdar}, \&
  {Straka}}]{Demarque08}
{Demarque}, P., {Guenther}, D.B., {Li}, L.H., {Mazumdar}, A. and {Straka},
  C.W., 2008, \apss, 316, 31

\bibitem[{{Dotter} {et~al.}(2008){Dotter}, {Chaboyer}, {Jevremovi{\'c}},
  {Kostov}, {Baron}, \& {Ferguson}}]{Dotter08}
{Dotter}, A., {Chaboyer}, B., {Jevremovi{\'c}}, D., {Kostov}, V., {Baron}, E.
  et~al, 2008, \apjs, 178, 89

\bibitem[{{Ekstr{\"o}m} {et~al.}(2012){Ekstr{\"o}m}, {Georgy}, {Eggenberger},
  {Meynet}, {Mowlavi}, {Wyttenbach}, {Granada}, {Decressin}, {Hirschi},
  {Frischknecht}, {Charbonnel}, \& {Maeder}}]{Ekstrom12}
{Ekstr{\"o}m}, S., {Georgy}, C., {Eggenberger}, P., {Meynet}, G., {Mowlavi}, N.
  et~al, 2012, \aap, 537, A146

\bibitem[{{Feiden} \& {Chaboyer}(2014)}]{Feiden14}
{Feiden}, G.A. and {Chaboyer}, B., 2014, \apj, 789, 53

\bibitem[{{Feiden} {et~al.}(2015){Feiden}, {Jones}, \& {Chaboyer}}]{Feiden15}
{Feiden}, G.A., {Jones}, J. and {Chaboyer}, B., 2015, in Cambridge Workshop on
  Cool Stars, Stellar Systems, and the Sun, Vol.~18, 18th Cambridge Workshop on
  Cool Stars, Stellar Systems, and the Sun, {van Belle}, G.T. and {Harris},
  H.C., eds., p. 171

\bibitem[{{Gennaro} {et~al.}(2012){Gennaro}, {Prada Moroni}, \&
  {Tognelli}}]{Gennaro12}
{Gennaro}, M., {Prada Moroni}, P.G. and {Tognelli}, E., 2012, \mnras, 420, 986

\bibitem[{{Herbst}(2008)}]{Herbst08}
{Herbst}, W., 2008, {Handbook of Star Forming Regions, Volume I: The Northern
  Sky}, {Reipurth}, B., ed., p. 372

\bibitem[{{Hillenbrand} \& {White}(2004)}]{Hillenbrand04}
{Hillenbrand}, L.A. and {White}, R.J., 2004, \apj, 604, 741

\bibitem[{{Jackson} \& {Jeffries}(2014)}]{Jackson14b}
{Jackson}, R.J. and {Jeffries}, R.D., 2014, \mnras, 445, 4306

\bibitem[{{Jackson} {et~al.}(2016){Jackson}, {Jeffries}, {Randich},
  {Bragaglia}, {Carraro}, {Costado}, {Flaccomio}, {Lanzafame}, {Lardo},
  {Monaco}, {Morbidelli}, {Smiljanic}, \& {Zaggia}}]{Jackson16}
{Jackson}, R.J., {Jeffries}, R.D., {Randich}, S., {Bragaglia}, A., {Carraro},
  G. et~al, 2016, \aap, 586, A52

\bibitem[{{Jeffries} {et~al.}(2014){Jeffries}, {Jackson}, {Cottaar}, {Koposov},
  {Lanzafame}, {Meyer}, {Prisinzano}, {Randich}, {Sacco}, {Brugaletta},
  {Caramazza}, {Damiani}, {Franciosini}, {Frasca}, {Gilmore}, {Feltzing},
  {Micela}, {Alfaro}, {Bensby}, {Pancino}, {Recio-Blanco}, {de Laverny},
  {Lewis}, {Magrini}, {Morbidelli}, {Costado}, {Jofr{\'e}}, {Klutsch}, {Lind},
  \& {Maiorca}}]{Jeffries14}
{Jeffries}, R.D., {Jackson}, R.J., {Cottaar}, M., {Koposov}, S.E., {Lanzafame},
  A.C. et~al, 2014, \aap, 563, A94

\bibitem[{{Johnson} \& {Morgan}(1953)}]{Johnson53}
{Johnson}, H.L. and {Morgan}, W.W., 1953, \apj, 117, 313

\bibitem[{{Kraus} {et~al.}(2015){Kraus}, {Cody}, {Covey}, {Rizzuto}, {Mann}, \&
  {Ireland}}]{Kraus15}
{Kraus}, A.L., {Cody}, A.M., {Covey}, K.R., {Rizzuto}, A.C., {Mann}, A.W.
  et~al, 2015, \apj, 807, 3

\bibitem[{{Kraus} {et~al.}(2011){Kraus}, {Tucker}, {Thompson}, {Craine}, \&
  {Hillenbrand}}]{Kraus11}
{Kraus}, A.L., {Tucker}, R.A., {Thompson}, M.I., {Craine}, E.R. and
  {Hillenbrand}, L.A., 2011, \apj, 728, 48

\bibitem[{{Littlefair} {et~al.}(2010){Littlefair}, {Naylor}, {Mayne},
  {Saunders}, \& {Jeffries}}]{Littlefair10}
{Littlefair}, S.P., {Naylor}, T., {Mayne}, N.J., {Saunders}, E.S. and
  {Jeffries}, R.D., 2010, \mnras, 403, 545

\bibitem[{{Mamajek} \& {Bell}(2014)}]{Mamajek14}
{Mamajek}, E.E. and {Bell}, C.P.M., 2014, \mnras, 445, 2169

\bibitem[{{Mathieu} {et~al.}(2007){Mathieu}, {Baraffe}, {Simon}, {Stassun}, \&
  {White}}]{Mathieu07}
{Mathieu}, R.D., {Baraffe}, I., {Simon}, M., {Stassun}, K.G. and {White}, R.,
  2007, Protostars and Planets V, 411

\bibitem[{{Melis} {et~al.}(2014){Melis}, {Reid}, {Mioduszewski}, {Stauffer}, \&
  {Bower}}]{Melis14}
{Melis}, C., {Reid}, M.J., {Mioduszewski}, A.J., {Stauffer}, J.R. and {Bower},
  G.C., 2014, Science, 345, 1029

\bibitem[{{Meynet} \& {Maeder}(2000)}]{Meynet00}
{Meynet}, G. and {Maeder}, A., 2000, \aap, 361, 101

\bibitem[{{Naylor}(2009)}]{Naylor09}
{Naylor}, T., 2009, \mnras, 399, 432

\bibitem[{{Pecaut} \& {Mamajek}(2016)}]{Pecaut16}
{Pecaut}, M.J. and {Mamajek}, E.E., 2016, \mnras, 461, 794

\bibitem[{{Raghavan} {et~al.}(2010){Raghavan}, {McAlister}, {Henry}, {Latham},
  {Marcy}, {Mason}, {Gies}, {White}, \& {ten Brummelaar}}]{Raghavan10}
{Raghavan}, D., {McAlister}, H.A., {Henry}, T.J., {Latham}, D.W., {Marcy}, G.W.
  et~al, 2010, \apjs, 190, 1

\bibitem[{{Soderblom} {et~al.}(2014){Soderblom}, {Hillenbrand}, {Jeffries},
  {Mamajek}, \& {Naylor}}]{Soderblom14}
{Soderblom}, D.R., {Hillenbrand}, L.A., {Jeffries}, R.D., {Mamajek}, E.E. and
  {Naylor}, T., 2014, Protostars and Planets VI, 219

\bibitem[{{Soderblom} {et~al.}(2005){Soderblom}, {Nelan}, {Benedict},
  {McArthur}, {Ramirez}, {Spiesman}, \& {Jones}}]{Soderblom05}
{Soderblom}, D.R., {Nelan}, E., {Benedict}, G.F., {McArthur}, B., {Ramirez}, I.
  et~al, 2005, \aj, 129, 1616

\bibitem[{{Somers} \& {Pinsonneault}(2015)}]{Somers15}
{Somers}, G. and {Pinsonneault}, M.H., 2015, \apj, 807, 174

\bibitem[{{Stassun} {et~al.}(2014){Stassun}, {Feiden}, \& {Torres}}]{Stassun14}
{Stassun}, K.G., {Feiden}, G.A. and {Torres}, G., 2014, \nar, 60, 1

\bibitem[{{Stauffer} {et~al.}(2007){Stauffer}, {Hartmann}, {Fazio}, {Allen},
  {Patten}, {Lowrance}, {Hurt}, {Rebull}, {Cutri}, {Ramirez}, {Young}, {Rieke},
  {Gorlova}, {Muzerolle}, {Slesnick}, \& {Skrutskie}}]{Stauffer07}
{Stauffer}, J.R., {Hartmann}, L.W., {Fazio}, G.G., {Allen}, L.E., {Patten},
  B.M. et~al, 2007, \apjs, 172, 663

\bibitem[{{Stauffer} {et~al.}(1998){Stauffer}, {Schild}, {Barrado y Navascues},
  {Backman}, {Angelova}, {Kirkpatrick}, {Hambly}, \& {Vanzi}}]{Stauffer98a}
{Stauffer}, J.R., {Schild}, R., {Barrado y Navascues}, D., {Backman}, D.E.,
  {Angelova}, A.M. et~al, 1998, \apj, 504, 805

\bibitem[{{Tognelli} {et~al.}(2015){Tognelli}, {Prada Moroni}, \&
  {Degl'Innocenti}}]{Tognelli15}
{Tognelli}, E., {Prada Moroni}, P.G. and {Degl'Innocenti}, S., 2015, \mnras,
  449, 3741

\end{thebibliography}

\end{document}